# Comments on the "Proof of the atmospheric greenhouse effect" by Arthur P. Smith

**Gerhard Kramm**[1], **Ralph Dlugi**[2], and **Michael Zelger**[2]

[1]University of Alaska Fairbanks, Geophysical Institute
903 Koyukuk Drive, P.O. Box 757320, Fairbanks, AK 99775-7320, USA

[2]Arbeitsgruppe Atmosphärische Prozesse (AGAP),
Gernotstraße, D-80804 Munich, Germany

**Abstract:** In this paper it is shown that Smith (2008) used inappropriate and inconsistent formulations in averaging various quantities over the entire surface of the Earth considered as a sphere. Using two instances of averaging procedures as customarily applied in studies on turbulence, it is shown that Smith's formulations are highly awkward. Furthermore, Smith's discussion of the infrared absorption in the atmosphere is scrutinized and evaluated. It is shown that his attempt to refute the criticism of Gerlich and Tscheuschner (2007, 2009) on the so-called greenhouse effect is rather fruitless.

1. Introduction

The so-called greenhouse effect of the atmosphere is commonly explained as followed (see Glossary of Meteorology, American Meteorological Society, http://amsglossary.allenpress.com/glossary/search?id=greenhouse-effect1):

"The heating effect exerted by the atmosphere upon the Earth because certain trace gases in the atmosphere (water vapor, carbon dioxide, etc.) absorb and reemit infrared radiation.

Most of the sunlight incident on the Earth is transmitted through the atmosphere and absorbed at the Earth's surface. The surface tries to maintain energy balance in part by emitting its own radiation, which is primarily at the infrared wavelengths characteristic of the Earth's temperature. Most of the heat radiated by the surface is absorbed by trace gases in the overlying atmosphere and reemitted in all directions. The component that is radiated downward warms the Earth's surface more than would occur if only the direct sunlight were absorbed. The magnitude of this enhanced warming is the greenhouse effect. Earth's annual mean surface temperature of 15°C is 33°C higher as a result of the greenhouse effect than the mean temperature resulting from radiative equilibrium of a blackbody at the Earth's mean distance from the sun. The term "greenhouse effect" is something of a misnomer. It is an analogy to the trapping of heat by the glass panes of a greenhouse, which let sunlight in. In the atmosphere, however, heat is trapped radiatively, while in an actual greenhouse, heat is mechanically prevented from escaping (via convection) by the glass enclosure."

According to this explanation we may carry out the following "thought experiment" of radiative equilibrium also called a zero-dimensional model, where we assume an Earth without an atmosphere. A consequence of this assumption is that (a) absorption of solar and terrestrial (infrared) radiation, (b) scattering of solar radiation by molecules and particulate matter, (c) radiative emission of energy in the infrared range, (d) convection and advection of heat, and (e) phase transition processes related to the formation and depletion of clouds can be ignored. The incoming flux of solar radiation, $F_{S\downarrow}$, that is absorbed at the Earth's surface is given by (see Figure 1 and section 4)

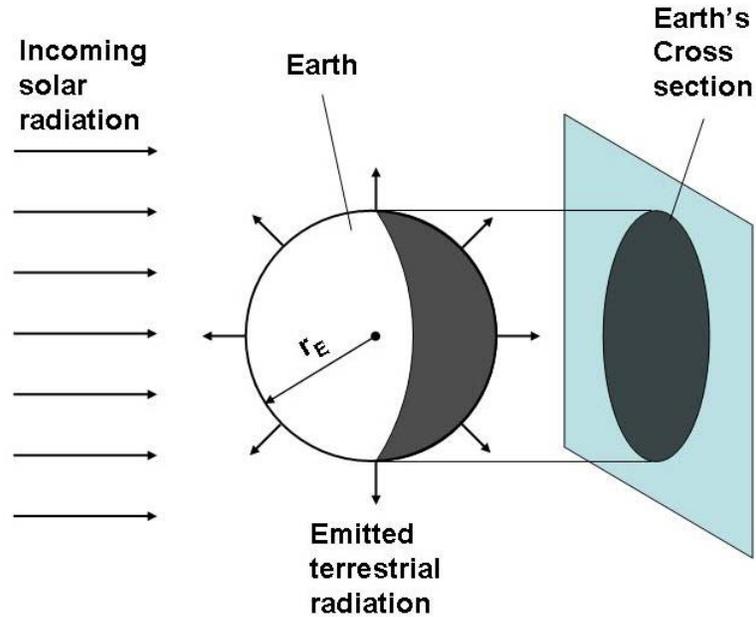

**Figure 1:** Sketch of the planetary radiative equilibrium (adopted from Petty, 2004).

$$F_{S\downarrow} = \pi\, r_E^{\,2}\, S\,(1 - \alpha_E) \quad . \tag{1.1}$$

Here, S is the solar constant, i.e., the solar irradiance at the top of the atmosphere calculated for a mean distance (1 Astronomic Unit = AU) between the sun and the Earth of $d_0 \cong 1.496 \cdot 10^8$ km (e.g., Iqbal, 1983; Vardavas and Taylor, 2007). Traditionally, a value for the solar constant close to $S \cong 1367\ \text{W m}^{-2}$ is recommended (e.g., Liou, 2002; Petty, 2004, Bohren and Clothiaux, 2006), but, as illustrated in Figure 2, the value obtained from recent satellite observations using TIM (Total Irradiance Monitoring; launched in 2003) is close to $S \cong 1361\ \text{W m}^{-2}$. The basis for this modified value is a more reliable, improved absolute calibration. Furthermore, $r_E \cong 6,371$ km is the mean radius of the Earth considered as a sphere, and $\alpha_E \approx 0.30$ is the planetary albedo of the Earth. This value is based on satellite observations of the real Earth enveloped by its atmosphere. The planetary albedo of the Earth of our thought experiment might be different.

If we assume that the temperature, $T_e$, of the Earth's surface is uniform, i.e., it is assumed that it depends neither on the longitude nor on the latitude, the total flux of infrared radiation emitted by the Earth's surface, $F_{IR\uparrow}$, as a function of this temperature and the planetary emissivity, $\varepsilon_E$, will be given by (Figure 1 and section 4)

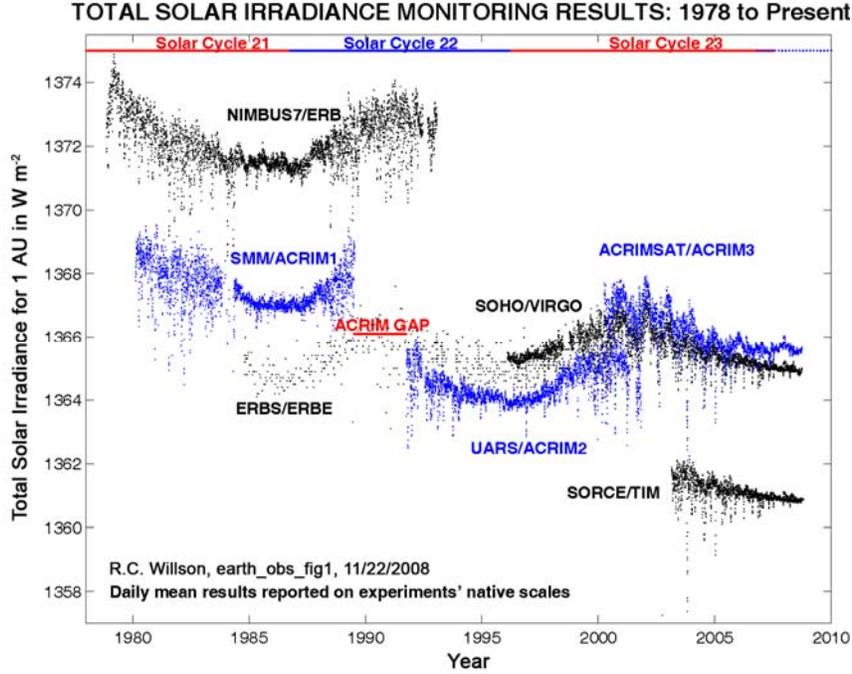

**Figure 2:** Satellite observations of total solar irradiance. It comprises of the observations of seven independent experiments: (a) Nimbus7/Earth Radiation Budget experiment (1978 - 1993), (b) Solar Maximum Mission/Active Cavity Radiometer Irradiance Monitor 1 (1980 - 1989), (c) Earth Radiation Budget Satellite/Earth Radiation Budget Experiment (1984 - 1999), (d) Upper Atmosphere Research Satellite/Active cavity Radiometer Irradiance Monitor 2 (1991 - 2001), (e) Solar and Heliospheric Observer/Variability of solar Irradiance and Gravity Oscillations (launched in 1996), (f) ACRIM Satellite/Active cavity Radiometer Irradiance Monitor 3 (launched in 2000), and (g) Solar Radiation and Climate Experiment/Total Irradiance Monitor (launched in 2003). The figure is based on Dr. Richard C. Willson's earth_obs_fig1, updated on November 22, 2008 (see http://www.acrim.com/).

$$F_{IR\uparrow} = 4 \pi r_E^2 \varepsilon_E \sigma T_e^4 \quad . \tag{1.2}$$

This equation is based on the power law of Stefan (1879) and Boltzmann (1884), where $\sigma \cong 5.67 \cdot 10^{-8}$ W m$^{-2}$ K$^{-4}$ is Stefan's constant. If we assume that there is a so-called planetary radiative equilibrium, i.e., $F_{S\downarrow} = F_{IR\uparrow}$ (see Appendix), we will obtain (e.g., Hansen et al., 1984; Hartmann, 1994; Liou, 2002; Petty, 2004; Vardavas and Taylor, 2007)

$$S(1 - \alpha_E) = 4 \varepsilon_E \sigma T_e^4 \quad . \tag{1.3}$$

This equation characterizes the energy balance of this simplified system. Thus, the temperature $T_e$ can be expressed by

$$T_e = \left( \frac{(1-\alpha_E)S}{4\,\varepsilon_E\,\sigma} \right)^{\frac{1}{4}} . \qquad (1.4)$$

Assuming that the Earth is a black body ($\varepsilon_E = 1$), and using the values mentioned above yields then $T_e \approx 255\,\text{K}$. Since the global average of temperatures observed in the close vicinity of the Earth's surface corresponds to $\langle T \rangle \approx 288\,\text{K}$, the difference between the mean global temperature and the temperature of the planetary radiative equilibrium amounts to $\Delta T = \langle T \rangle - T_e \approx 33\,\text{K}$. Therefore, as stated in the Glossary Of Meteorology of the American Meteorological Society, the so-called greenhouse effect of the atmosphere causes a temperature increase of about $33\,\text{K}$, regardless of the fact that the atmosphere is an open thermodynamic system in which various processes listed before may take place, but not a simple "greenhouse" that causes the trapping of radiative heat. Gerlich and Tscheuschner (2007, 2009), therefore, fiercely attacked this kind of simplification. They stated that (a) there are no common physical laws between the warming phenomenon in glass houses and the fictitious atmospheric greenhouse effect, (b) there are no calculations to determine an average surface temperature of a planet, (c) the frequently mentioned difference of 33 K is a meaningless number calculated wrongly, (d) the formulae of cavity radiation are used inappropriately, (e) the assumption of a radiative balance is unphysical, (f) thermal conductivity and friction must not be set to zero, the atmospheric greenhouse conjecture is falsified.

In his paper entitled "Proof of the atmospheric greenhouse effect" Smith (2008) tried to refute the criticism of Gerlich and Tscheuschner on the so-called greenhouse effect. In doing so, he introduced effective quantities by averaging over the entire globe (see his Eqs. (7) to (9)). Obviously, there is a violation of basic rules of calculus. This fact is explained in section 2. In section 3 two examples of averaging principles used in studies on time dependent processes like turbulence are presented. Based on these examples it is shown the Smith's planetary averages are affected by some inconsistent formulations. In addition, in section 4 it is shown that Smith's use of, at least, three different planetary averages are not necessary to derive the respective equations of the planetary radiation balance for an Earth without an atmosphere. In section 5 Smith's discussion of the infrared absorption in the atmosphere is scrutinized using the two-layer model with the common assumption of radiative equilibrium.

## 2. The planetary-averaging procedure

If we considers the true shape of the Earth (the radius of the equator, 6,378 km, is larger than the radius to the poles, 6,356 km, owing to centrifugal forces), the average over the entire Earth's surface is given by (e.g., Riley et al., 1998, pp.142-143)

$$\langle \Psi \rangle = \frac{\int_\Omega \Psi(r,\theta,\phi)\,r^2(\theta,\phi)\,d\Omega}{\int_\Omega r^2(\theta,\phi)\,d\Omega} . \qquad (2.1)$$

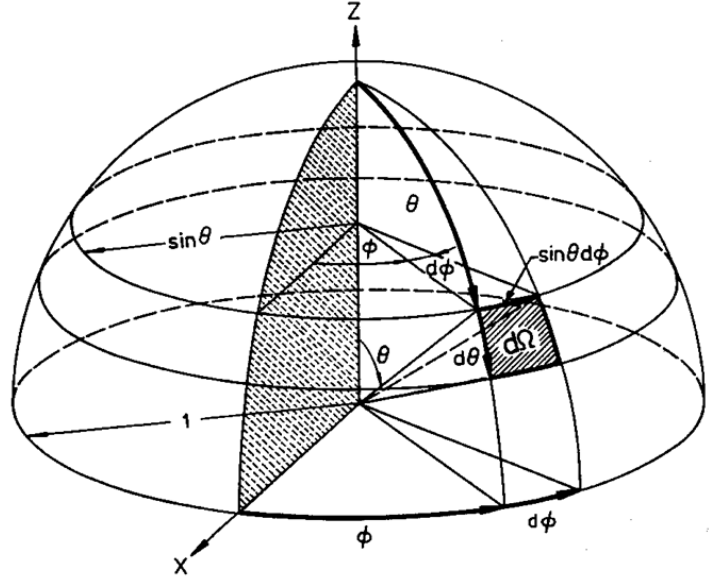

**Figure 3:** Mathematical representation of the solid angle (adopted from Kasten and Raschke, 1974). Here, $d\Omega = \sin\theta\, d\theta\, d\phi$ is the differential solid angle, where $\theta$ and $\phi$ are the zenith and azimuthal angles, respectively.

Here, $\Psi(r,\theta,\phi)$ is an arbitrary variable, $r(\theta,\phi)$ is the radius, $\Omega$ is the solid angle and $d\Omega = \sin\theta\, d\theta\, d\phi$ is the differential solid angle, where $\theta$ and $\phi$ are the zenith and azimuthal angles, respectively, of a spherical coordinate frame (see Figure 3). Since Smith only considered a sphere, i.e., $r(\theta,\phi) = r = \text{const.}$ and $\Psi(r,\theta,\phi) = \Psi(\theta,\phi)$, Eq. (2.1) results in

$$\langle \Psi \rangle = \frac{1}{4\pi} \int_\Omega \Psi(\theta,\phi)\, d\Omega \quad , \tag{2.2}$$

where the solid angle of a sphere is given by $\Omega = 4\pi$. Note that the radius of the sphere does not play any further role. For $\Psi(\theta,\phi) = T^4(\theta,\phi)$, we immediately obtain

$$\langle T^4 \rangle = \frac{1}{4\pi} \int_\Omega T^4(\theta,\phi)\, d\Omega \quad . \tag{2.3}$$

For the purpose of comparison: Eq. (7) of Smith reads

$$T_{\text{eff}}(t)^4 = \frac{1}{4\pi r^2} \int T(\mathbf{x},t)^4\, d\mathbf{x} \quad . \tag{2.4}$$

He defined the local surface temperature by $T(\mathbf{x},t)$ and the local emissivity by $\varepsilon(\mathbf{x},t)$, where $t$ is time and $\mathbf{x}$ is the position vector. To still use the radius in this equation is awkward because, as stated before, the radius plays no role if a sphere is considered. According to Smith, $T_{eff}(t)^4$ is defined as an average over the planetary surface. Note that even this notation is awkward because the quantity planetary-averaged is $T(\mathbf{x},t)^4$, i.e., it should write $\left(T(t)^4\right)_{eff} \neq T_{eff}(t)^4$. Since the position vector has the physical unit of a length, it is obvious that Smith's equations (3), (6) to (8), and (11) are flawed by dimensional inaccuracies. The left-hand side of Smith's Eq. (7) listed here as (2.4), for instance, demands $K^4$. Whereas the right-hand side of that equations offers $K^4 \, m^{-1}$.

For the radiative emission of energy, $\Psi(\theta, \phi) = E(\theta, \phi) = \varepsilon(\theta, \phi) \, \sigma \, T^4(\theta, \phi)$, Eq. (2.2) provides

$$\langle E \rangle = \frac{1}{4\pi} \int_\Omega E(\theta, \phi) \, d\Omega \tag{2.5}$$

or

$$\langle \varepsilon \, T^4 \rangle = \frac{1}{4\pi} \int_\Omega \varepsilon(\theta, \phi) \, T^4(\theta, \phi) \, d\Omega = \frac{1}{4\pi} \int_\Omega \varepsilon(\theta, \phi) \, T^4(\theta, \phi) \sin\theta \, d\theta \, d\phi \tag{2.6}$$

Note, that Eq. (2.5) or (2.6) are formulated for the product term in the Stefan – Boltzmann Law which cannot simply be separated from each other. Multiplying with $\sigma$, gives the energy flux densities for each part of the Earth's surface element. For such conserved quantities balance equations are valid and averaging procedures like Eq. (2.1) or (2.5) can be applied, because such quantities are additive. Temperature is an observed property, but has to be transformed into energy quantities before averaging. For the purpose of comparison: Equation (8) of Smith reads

$$\varepsilon_{eff}(t) = \frac{1}{4\pi r^2 \, T_{eff}(t)^4} \int \varepsilon(\mathbf{x},t) \, T(\mathbf{x},t)^4 \, d\mathbf{x} \tag{2.7}$$

Smith claimed that this equation should be considered as a definition of $\varepsilon_{eff}(t)$. If this is true, then this quantity cannot be considered as an average over the planetary surface as he stated in the sentence directly followed after his Eq. (6)[1].

If we set $\Psi(\theta, \phi) = \varepsilon(\theta, \phi)$, the true planetary average of the emissivity is given by

$$\langle \varepsilon \rangle = \frac{1}{4\pi} \int_\Omega \varepsilon(\theta, \phi) \, d\Omega \quad , \tag{2.8}$$

---

[1] Smith wrote: "Similar to the effective albedo, an effective emissivity and effective radiative temperature can be defined as averages over the planetary surface:"

or in Smith's notation,

$$\varepsilon_{\text{eff}}(t) = \frac{1}{4\pi r^2} \int \varepsilon(\mathbf{x}, t) \, d\mathbf{x} \quad . \tag{2.9}$$

This means that the planetary-averaged emissivity is **not** given by

$$\langle \varepsilon \rangle = \frac{1}{4\pi \langle T^4 \rangle} \int_\Omega \varepsilon(\theta, \phi) \, T^4(\theta, \phi) \, d\Omega \tag{2.10}$$

because it would mean that $\langle \varepsilon T^4 \rangle = \langle \varepsilon \rangle \langle T^4 \rangle$. The correct result is

$$\langle \varepsilon T^4 \rangle = \frac{1}{4\pi} \int_\Omega \varepsilon(\theta, \phi) \, T^4(\theta, \phi) \, d\Omega \neq \langle \varepsilon \rangle \langle T^4 \rangle = \frac{1}{4\pi} \int_\Omega \varepsilon(\theta, \phi) \, d\Omega \, \frac{1}{4\pi} \int_\Omega T^4(\theta, \phi) \, d\Omega \quad . \tag{2.11}$$

Dividing Eq. (2.6) by $\langle T^4 \rangle$ provides

$$\{\varepsilon\} = \frac{\langle \varepsilon T^4 \rangle}{\langle T^4 \rangle} = \frac{\frac{1}{4\pi} \int_\Omega \varepsilon(\theta, \phi) \, T^4(\theta, \phi) \, d\Omega}{\frac{1}{4\pi} \int_\Omega T^4(\theta, \phi) \, d\Omega} = \frac{\int_\Omega \varepsilon(\theta, \phi) \, T^4(\theta, \phi) \, d\Omega}{\int_\Omega T^4(\theta, \phi) \, d\Omega} \quad . \tag{2.12}$$

Here, $\{\varepsilon\}$ may be considered as a weighted average, i.e., it is the exact definition of a weighted emissivity. For the purpose of comparison: Using Smith's notation yields

$$\{\varepsilon(t)\} = \frac{\left(\varepsilon(t) \, T(t)^4\right)_{\text{eff}}}{T_{\text{eff}}(t)^4} = \frac{\frac{1}{4\pi r^2} \int \varepsilon(\mathbf{x}, t) \, T(\mathbf{x}, t)^4 \, d\mathbf{x}}{\frac{1}{4\pi r^2} \int T(\mathbf{x}, t)^4 \, d\mathbf{x}} = \frac{\int \varepsilon(\mathbf{x}, t) \, T(\mathbf{x}, t)^4 \, d\mathbf{x}}{\int T(\mathbf{x}, t)^4 \, d\mathbf{x}} \neq \frac{1}{4\pi r^2} \int \varepsilon(\mathbf{x}, t) \, d\mathbf{x}$$

$$\tag{2.13}$$

i.e, $\{\varepsilon(t)\} \neq \varepsilon_{\text{eff}}(t)$. This means that Smith's formulations are affected by improper definitions of effective and planetary-averaged quantities.

In his Eq. (11) Smith also defines the planetary average of the temperature by (using his notation)

$$T_{\text{ave}}(t) = \frac{1}{4\pi r^2} \int T(\mathbf{x}, t) \, d\mathbf{x} \quad . \tag{2.14}$$

Besides the fact that he still included the radius even though he considered a sphere, this definition is in agreement with that reflected by Eq. (2.2). However, using, at least, three different averages (see Eqs. (7) to (9) and (11) of Smith) makes no sense because the respective calculations can be performed using the definition given by Eq. (2.1), or in the case of a sphere, by Eq. (2.2).

## 3. Excursion: Averaging procedures applied in studies on turbulence

The following two examples serve as guidance how averages and deviation from these values have to be treated consistently.

### 3.1 Time averages

Many time dependent processes like turbulence are stochastic. A stochastic process is called ergodic when the ensemble averages may be computed as time averages (e.g., Liepmann, 1952). The time average of an arbitrary time-dependent quantity $f(t)$ is given by

$$\bar{f} = \lim_{T \to \infty} \frac{1}{T} \int_{t_0}^{t_0+T} f(t)\, dt \quad , \tag{3.1}$$

If we express $f(t)$ by $f(t) = \bar{f} + f'(t)$ the definition (3.1) yields

$$\bar{f} = \lim_{T \to \infty} \frac{1}{T} \int_{t_0}^{t_0+T} \left(\bar{f} + f'(t)\right) dt = \bar{f} + \lim_{T \to \infty} \frac{1}{T} \int_{t_0}^{t_0+T} f'(t)\, dt \quad . \tag{3.2}$$

From this equation one can infer that

$$\overline{f'} = \lim_{T \to \infty} \frac{1}{T} \int_{t_0}^{t_0+T} \left(f'(t)\right) dt = 0 \quad . \tag{3.3}$$

According to these equations the time average of the product of two arbitrary time-dependent quantities $f(t)$ and $g(t)$ is given by

$$\overline{f\,g} = \lim_{T \to \infty} \frac{1}{T} \int_{t_0}^{t_0+T} f(t)\, g(t)\, dt \quad . \tag{3.4}$$

Expressing $f(t) = \bar{f} + f'(t)$ and $g(t) = \bar{g} + g'(t)$ leads to

$$\overline{f\,g} = \lim_{T \to \infty} \frac{1}{T} \int_{t_0}^{t_0+T} \left(\bar{f} + f'(t)\right)\left(\bar{g} + g'(t)\right) dt \tag{3.5}$$

or

$$\overline{fg} = \lim_{T \to \infty} \frac{1}{T} \int_{t_0}^{t_0+T} \left( \overline{f}\,\overline{g} + \overline{g}\,f'(t) + \overline{f}\,g'(t) + f'(t)g'(t) \right) dt \tag{3.6}$$

or

$$\overline{fg} = \overline{f}\,\overline{g} + \lim_{T \to \infty} \frac{\overline{g}}{T} \int_{t_0}^{t_0+T} f'(t)\,dt + \lim_{T \to \infty} \frac{\overline{f}}{T} \int_{t_0}^{t_0+T} g'(t)\,dt + \lim_{T \to \infty} \frac{1}{T} \int_{t_0}^{t_0+T} (f'(t)g'(t))\,dt \quad . \tag{3.7}$$

According to Eq. (3.3) the second term and the third term on the right-hand side of this equation are equal to zero. Thus, we obtain

$$\overline{fg} = \overline{f}\,\overline{g} + \lim_{T \to \infty} \frac{1}{T} \int_{t_0}^{t_0+T} (f'(t)g'(t))\,dt \tag{3.8}$$

or

$$\overline{fg} = \overline{f}\,\overline{g} + \overline{f'g'} \tag{3.9}$$

with

$$\overline{f'g'} = \lim_{T \to \infty} \frac{1}{T} \int_{t_0}^{t_0+T} (f'(t)g'(t))\,dt \tag{3.10}$$

If $\overline{f'g'}$ could be expressed by $\overline{f'g'} = \overline{f'}\,\overline{g'}$ or

$$\lim_{T \to \infty} \frac{1}{T} \int_{t_0}^{t_0+T} (f'(t)g'(t))\,dt = \lim_{T \to \infty} \frac{1}{T} \int_{t_0}^{t_0+T} f'(t)\,dt \; \lim_{T \to \infty} \frac{1}{T} \int_{t_0}^{t_0+T} g'(t)\,dt \quad , \tag{3.11}$$

as suggested by Smith with his equation (8), we would have

$$\overline{f'g'} = 0 \quad . \tag{3.12}$$

Consequently, all variance or covariance terms that occur in the governing equations of turbulent systems would be equal to zero. This, of course, is not true.

### 3.2   Density-weighted averages

It is well known that conventional Reynolds averaging similar to Eq. (3.1) leads to various shortcomings in the set of governing equations for turbulent compressible systems, even though this averaging technique can accurately be performed. If we ignore density fluctuation terms, as customarily done within the framework of the Boussinesq approximation (except in those terms expressing the effect of the gravity field on the density fluctuations), the possibility to describe physical processes as a whole will clearly be restricted, as already pointed out by Montgomery (1954) and Fortak (1969). As stated by Kramm and Meixner (2000), the key questions that still remain are (a) how to average the governing macroscopic equations in the case of turbulent atmospheric flows, and (b) what are the consequences of such an averaging, not only for atmospheric trace species, but also for total mass, momentum, and various energy forms.

As argued by Herbert (1975), Pichler (1984), Cox (1995), Kramm et al. (1995), Thomson (1995) and Venkatram (1998), the density-weighted averaging procedure suggested by Hesselberg (1926) is well appropriate to formulate the balance equation of momentum, internal energy (alternatively enthalpy), and gaseous and particulate matter. In the following we summarize some advantages of the application of a physically more appropriate other density weighted average. Hesselberg's averaging calculus is based on (e.g., van Mieghem, 1949, 1973; Herbert, 1975; Kramm and Meixner, 2000)

$$\hat{\varphi} = \hat{\varphi}(\mathbf{r}) = \frac{\int_G \rho(\mathbf{r}, \mathbf{r}') \varphi(\mathbf{r}, \mathbf{r}') \, dG'}{\int_G \rho(\mathbf{r}, \mathbf{r}') \, dG'} = \frac{\overline{\rho \varphi}}{\overline{\rho}} \quad , \tag{3.13}$$

where $\rho$ is the air density, and $\varphi(\mathbf{r})$ is a specific quantity like the wind vector, $\mathbf{v}(\mathbf{r})$, the specific internal energy, $e(\mathbf{r})$, the temperature, $T(\mathbf{r})$, the specific humidity, $q(\mathbf{r})$, etc. Note that the overbar ($\overline{\phantom{x}}$) defines the conventional Reynolds mean; whereas the hat (ˆ) denotes Hesselberg's density-weighted average, and a double prime (") marks the departure from that. Thus, $\hat{\varphi}$ is the density-weighted average in space and time of $\varphi(\mathbf{r})$, and the fluctuation, $\varphi''(\mathbf{r})$, is the difference between the former and the latter. Furthermore, $\mathbf{r}$ is the four-dimensional vector of space and time in the original coordinate system, $\mathbf{r}'$ is that of the averaging domain G where its origin, $\mathbf{r}' = 0$, is assumed to be $\mathbf{r}$, and $dG' = d^3r' dt'$. The averaging domain G is given by $G = \int_G dG'$. Hence, the quantity $\hat{\varphi}$ represents the mean values of $\varphi(\mathbf{r})$ for the averaging domain G at the location $\mathbf{r}$. It is obvious that $\overline{\rho \varphi''} = 0$. Arithmetic rules can be found, for instance, in van Mieghem (1949, 1973), Herbert (1975), Pichler (1984), and Kramm et al. (1995). Obviously, Eqs. (2.12) and (3.13) are quite similar. The advantage of Hesselberg's density-weighted average can be summarized as follows:

(a) The equation of continuity (van Mieghem, 1949, 1973; Herbert, 1975; Pichler, 1984; Cox, 1995; Kramm et al., 1995),

$$\frac{\partial \overline{\rho}}{\partial t} + \nabla \cdot \left( \overline{\rho} \, \hat{\mathbf{v}} \right) = 0 \quad , \tag{3.14}$$

keeps its form. Here, $\nabla$ is the nabla operator.

(b) The mean value of kinetic energy can exactly be split into the kinetic energy of the mean motion and mean value of the kinetic energy of the eddying motion (van Mieghem, 1949, 1973; Pichler, 1984; Kramm et al., 1995), i.e.,

$$\frac{1}{2}\overline{\rho\,\mathbf{v}^2} = \frac{1}{2}\overline{\rho}\,\hat{\mathbf{v}}^2 + \frac{1}{2}\overline{\rho\,\mathbf{v}''^2} \quad . \tag{3.15}$$

(c) The substantial derivative with respect to time of any property, $d/dt$, can exactly be expressed by Euler's operator for the averaged turbulent flow of a compressible atmosphere called a Hesselberg fluid (Kramm and Meixner 2000),

$$\frac{d}{dt} = \frac{\partial}{\partial t} + \hat{\mathbf{v}} \cdot \nabla \quad . \tag{3.16}$$

Such a derivation is impossible in the case of conventional Reynolds averaging because an expression for Euler's operator similar to Eq. (3.16) can only be deduced when the density fluctuation terms are ignored.

The local balance equation for internal energy (first law of thermodynamics) for the turbulent atmosphere (e.g., Kramm and Meixner, 2000),

$$\overline{\rho}\frac{d\hat{e}}{dt} + \nabla \cdot (\overline{\mathbf{R}} + \overline{\mathbf{J}_h} + \overline{\mathbf{F}_h}) = -\overline{p}\,\nabla \cdot \hat{\mathbf{v}} + \overline{\mathbf{v}'' \cdot \nabla p} - \overline{\mathbf{J}} : \nabla\hat{\mathbf{v}} - \overline{\mathbf{J} : \nabla\mathbf{v}''} \quad , \tag{3.17}$$

may serve as an example how Hesselberg's density-weighted average can successfully be applied. Here, $\overline{\mathbf{J}_h} = -c_p\,\overline{\rho}\,\boldsymbol{\alpha}\cdot\nabla\hat{T}$ and $\overline{\mathbf{F}_h} = \overline{\rho\,\mathbf{v}''h''}$ are the mean molecular and turbulent flux densities of enthalpy, respectively, where $|\overline{\mathbf{J}_h}| \ll |\overline{\mathbf{F}_h}|$ is usually valid except for the immediate vicinity of rigid walls, and h is the specific enthalpy that is related to the specific internal energy by $h = e + p/\rho$ (p is the air pressure). Furthermore, $\overline{\mathbf{R}}$ represents the mean radiative flux density, $\boldsymbol{\alpha} = \alpha\,\mathbf{E}$ is the second-rank tensor of thermal diffusivities customarily considered to be isotropic, $-\overline{\mathbf{J}} : \nabla\hat{\mathbf{v}} > 0$ is the direct dissipation of kinetic energy and $-\overline{\mathbf{J} : \nabla\mathbf{v}''} > 0$ is the turbulent dissipation of kinetic energy (the colon represents the double-scalar product of the tensor algebra), where $\mathbf{J}$ denotes the Stokes stress tensor. Moreover, the term $\overline{\mathbf{v}'' \cdot \nabla p}$ corresponds to the work that has to be done along with or against pressure gradient forces. Thus, especially Archimedian effects become relevant that transfer sensible and latent heat. Obviously, $\overline{\mathbf{v}'' \cdot \nabla p}$ can be either positive or negative mainly depending on the thermal stratification of the atmosphere.

Based on these considerations it is obvious that the use of a density-weighted average is required by the necessity to obtain exact governing equations for turbulent atmosphere that describe the physical behavior in an adequate manner. It is not based of any arbitrary choice of a weighted average.

Hesselberg's density-weighted average can be related to the conventional average Reynolds by

$$\hat{\varphi} = \overline{\varphi} + \frac{\overline{\rho' \varphi'}}{\overline{\rho}} \quad . \tag{3.18}$$

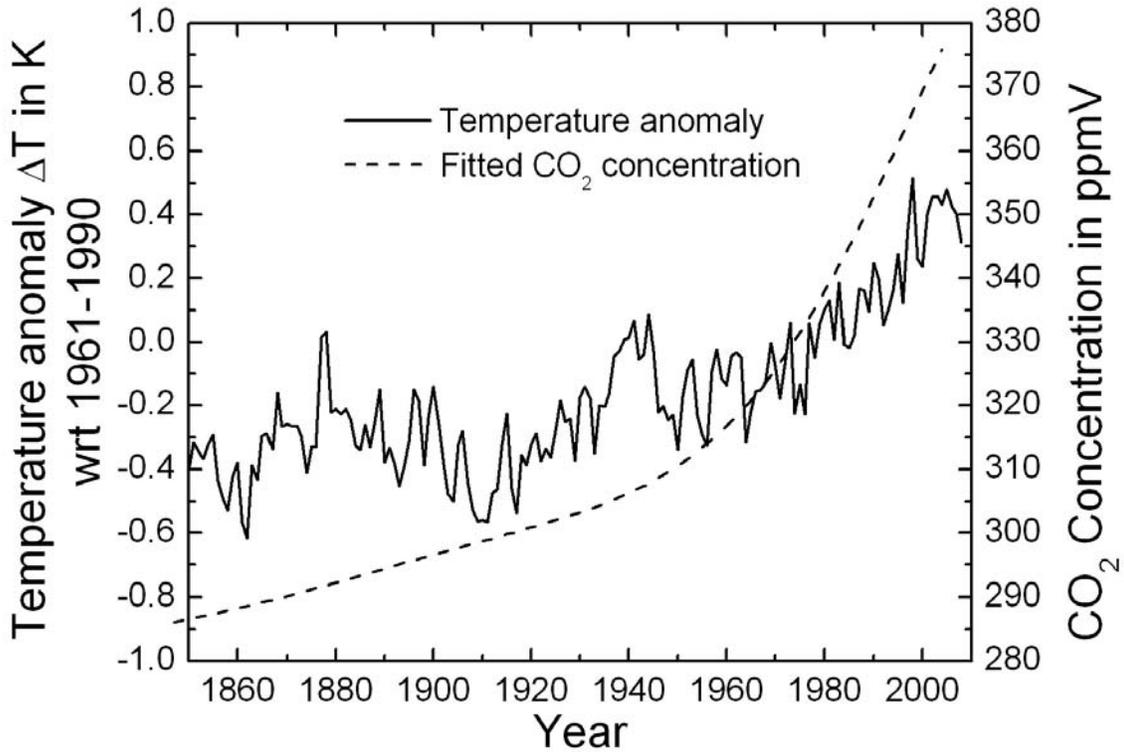

**Figure 4**: The increase of the globally averaged near-surface temperature (annual mean) expressed by the temperature anomaly with respect to the period 1961-1990. These temperature anomaly data were adopted from the Hadley Centre for Climate Prediction and Research, MetOffice, UK. A fit of the atmospheric carbon dioxide ($CO_2$) concentration adopted from Kramm et al. (2008) is also shown.

If on the right-hand side of this equation the amount of the first term is much larger than the amount of the second term the latter may be negligible and, hence, $\hat{\varphi} \approx \overline{\varphi}$. Such a simplification, of course, has to be justified on the basis of appropriate estimates. In accord with this expression, Eq. (2.12) has to be written as

$$\{\varepsilon\} = \langle \varepsilon \rangle + \frac{\langle \varepsilon * (T^4)* \rangle}{\langle T^4 \rangle} \quad , \tag{3.19}$$

where $\langle T^4 \rangle$ is given by Eq. (2.3) and $\langle \varepsilon \rangle$ by Eq. (2.8). This equation may also be used to express the term $\langle \varepsilon * (T^4)* \rangle$ by

$$\langle \varepsilon * (T^4)* \rangle = (\{\varepsilon\} - \langle \varepsilon \rangle) \langle T^4 \rangle \quad . \tag{3.20}$$

In the studies on turbulence Hesselberg's density-weighted average is highly favorable in formulating consistent governing equations. In the case of planetary averages the advantage of such formulations is highly questionable.

## 4.     The planetary radiation balance[2]

The total flux of solar radiation, $F_{S\downarrow}$, that is absorbed by the skin of the Earth being considered as of spherical shape and having no atmosphere is given by (see Appendix)

$$F_{S\downarrow} = r_E^2 \int_\Omega F(1 - \alpha(\Theta_0, \theta, \phi)) \cos \Theta_0 \, d\Omega = r_E^2 \int_0^{2\pi} \int_0^{\pi/2} F(1 - \alpha(\Theta_0, \theta, \phi)) \cos \Theta_0 \sin \theta \, d\theta \, d\phi \quad . \tag{4.1}$$

Here,

$$F = \left(\frac{r_S}{d}\right)^2 F_S \tag{4.2}$$

is the solar irradiance reaching the surface of the Earth, $r_S = 6.96 \cdot 10^6$ km is the visible radius of the Sun, $d$ is the time-dependent actual distance between the Sun and the Earth, and $F_S$ denotes the solar emittance (e.g., Liou, 2002). For the mean distance, $d_0$, $F$ becomes the solar constant $S$ so that we may write (e.g., Iqbal, 1983; Liou, 2002)

$$F = \left(\frac{d_0}{d}\right)^2 S \quad , \tag{4.3}$$

where the quantity $(d_0/d)^2$ is denoted here as the orbital effect. Since $(d_0/d)^2$ does not vary more than 3.5 percent (e.g., Liou, 2002), this orbital effect is usually ignored. Note, however, that the temperature difference, calculated using Eqs. (1.4) and (4.3), between perihelion and aphelion amounts to 4.2 K. This temperature difference is much larger than the increase of the globally averaged near-surface temperature during the last 160 years (see the temperature anomaly with respect to the period 1961-1990 illustrated in Figure 4). Moreover, $\alpha(\Theta_0, \theta, \phi)$ is the albedo that usually depends on $\Theta_0$, $\theta$ and $\phi$. The function $\cos \Theta_0$ is given by

$$\cos\Theta_0 = \sin\varphi\sin\delta + \cos\varphi\cos\delta\cos h \quad, \tag{4.4}$$

where $\Theta_0$ is the local zenith angle of the Sun's center, $\varphi$ is the latitude, $\delta$ is the solar declination angle, and h is the hour angle from the local meridian (e.g., Iqbal, 1983; Liou, 2002). The solar declination angle can be determined using $\sin\delta = \sin\beta\sin\gamma$, where $\beta = 23°27'$ is the oblique angle of the Earth's axis, and $\gamma$ is the true longitude of the Earth counted counterclockwise from the vernal equinox (e.g., Iqbal, 1983; Liou, 2002). The latitude is related to the zenith angle by $\varphi = \pi/2 - \theta$ so that formula (4.4) may be written as $\cos\Theta_0 = \cos\theta\sin\delta + \sin\theta\cos\delta\cos h$. Note that $\theta$ is ranging from zero to $\pi$, $\delta$ from $23°27'$S (Tropic of Capricorn) to $23°27'$N (Tropic of Cancer), and h from $-H$ to $H$, where H represents the half-day, i.e., from sunrise to solar noon or solar noon to sunset. It can be derived from Eq. (4.4) by setting $\Theta_0 = \pi/2$ (invalid at the poles) leading to $\cos H = -\tan\varphi\tan\delta$ (e.g., Iqbal, 1983; Liou, 2002).

---

[2] The entire section 4 is mainly based on section 3 of Kramm and Dlugi (2009).

Equation (4.1) may also be written as

$$F_{S\downarrow} = r_E^2 \left(\frac{d_0}{d}\right)^2 S \left\{ \int_0^{2\pi}\int_0^{\pi/2} \cos\Theta_0 \sin\theta\, d\theta\, d\phi - \int_0^{2\pi}\int_0^{\pi/2} \alpha(\Theta_0, \theta, \phi)\cos\Theta_0 \sin\theta\, d\theta\, d\phi \right\} \quad. \tag{4.5}$$

The equation cannot generally be solved in an analytical manner. If we, however, assume $\delta = \pi/2$ we will obtain $\cos\Theta_0 = \cos\theta$, i.e., the rotation axis of the Earth would always be parallel to the incoming solar radiation. Note, however, that the value of $\delta$ exceeds the Tropic of Cancer by far. This choice of $\delta$ results in

$$\begin{aligned}F_{S\downarrow} &= r_E^2 S \left\{ \int_0^{2\pi}\int_0^{\pi/2} \cos\theta \sin\theta\, d\theta\, d\phi - \int_0^{2\pi}\int_0^{\pi/2} \alpha(\theta, \phi)\cos\theta \sin\theta\, d\theta\, d\phi \right\} \\ &= r_E^2 S \left\{ \pi - \int_0^{2\pi}\int_0^{\pi/2} \alpha(\theta, \phi)\cos\theta \sin\theta\, d\theta\, d\phi \right\}\end{aligned} \quad, \tag{4.6}$$

where $(d_0/d)^2 \approx 1$ has been considered. Assuming a constant albedo for the bright side of the planet, i.e., $\alpha(\theta, \phi) = \alpha_E$, provides Eq. (1.1).

Smith introduced an effective albedo (see his Eq. (5)). For the purpose of comparison, Eq. (4.6) may also be written as

$$F_{S\downarrow} = \pi r_E^2 S (1 - \alpha_{eff}) \quad. \tag{4.7}$$

Here, the effective albedo, $\alpha_{eff}$, is defined by

$$\alpha_{eff} = \frac{1}{\pi} \int_0^{2\pi}\int_0^{\pi/2} \alpha(\theta, \phi)\cos\theta \sin\theta\, d\theta\, d\phi \quad, \tag{4.8}$$

where, in accord with Eq. (4.6),

$$\pi = \int_0^{2\pi} \int_0^{\pi/2} \cos\theta \sin\theta \, d\theta \, d\phi \quad . \tag{4.9}$$

According to Eq. (2.2), averaging over the bright side of the planet would provide for any arbitrary quantity $\Psi(\theta, \phi)$

$$\langle \Psi \rangle = \frac{\int_0^{2\pi} \int_0^{\pi/2} \Psi(\theta, \phi) \sin\theta \, d\theta \, d\phi}{\int_0^{2\pi} \int_0^{\pi/2} \sin\theta \, d\theta \, d\phi} = \frac{1}{2\pi} \int_0^{2\pi} \int_0^{\pi/2} \Psi(\theta, \phi) \sin\theta \, d\theta \, d\phi \quad . \tag{4.10}$$

From this point of view it is obvious that Smith's effective albedo is not based on averaging over the bright side of the planet. Thus, Smith's statement that $T_{eff}$ and $\varepsilon_{eff}$ are defined as planetary averages, similar to $\alpha_{eff}$ (see also footnote 1), is, therefore, misleading.

The total flux of infrared radiation, $F_{IR\uparrow}$, is given by (see Appendix)

$$F_{IR\uparrow} = r_E^2 \int_\Omega E(\theta, \phi) \, d\Omega = r_E^2 \int_0^{2\pi} \int_0^{\pi} E(\theta, \phi) \sin\theta \, d\theta \, d\phi \quad , \tag{4.11}$$

where now $\Omega$ is the solid angle for the entire planet, and $E(\theta, \phi) = \varepsilon(\theta, \phi) \sigma T^4(\theta, \phi)$ is the locally emitted infrared radiation. Only such a formulation is in substantial agreement with the derivation of the power law of Stefan (1879) and Boltzmann (1884). This derivation is based on (a) the integration of Planck's (1901) blackbody radiation law, for instance, over all frequencies and (b) the isotropic emission of radiant energy by a small spot (like a hole in the opaque walls of a cavity) into the adjacent half space. The emissivity $\varepsilon(\theta, \phi)$ and the surface temperature $T(\theta, \phi)$ depend on both $\theta$ and $\phi$. Customarily, it is assumed that $E(\theta, \phi)$ is uniformly distributed and may be replaced by $E_E = \varepsilon_E \sigma T_e^4$, where, again, $\varepsilon_E$ denotes the planetary emissivity. This assumption leads to

$$F_{IR\uparrow} = r_E^2 \varepsilon_E \sigma T_e^4 \int_0^{2\pi} \int_0^{\pi} \sin\theta \, d\theta \, d\phi = 4\pi r_E^2 \varepsilon_E \sigma T_e^4 \quad . \tag{4.12}$$

It completely agrees with Eq. (1.2). Note that Eq. (2.6) may serve to introduce a planetary-averaged value for the right-hand side of Eq. (4.11), as suggested by Smith. Thus, we would obtain

$$F_{IR\uparrow} = \sigma r_E^2 \langle \varepsilon T^4 \rangle \quad , \tag{4.13}$$

where, as shown in section 2, $\langle \varepsilon\, T^4 \rangle \neq \langle \varepsilon \rangle \langle T^4 \rangle$. From a physical point of view the use of such a planetary average is more satisfied than the assumption that $E(\theta, \phi)$ is uniformly distributed. The planetary radiation balance would be given by

$$S(1-\alpha_E) = 4\sigma \langle \varepsilon\, T^4 \rangle \quad . \tag{4.14}$$

This equation disagrees with Eq. (1.3), and a formula like Eq. (1.4) cannot be justified. Note that the use of the weighted average given by $\{\varepsilon\} = \langle \varepsilon\, T^4 \rangle / \langle T^4 \rangle$ (see Eq. (2.12)) does not provide any advantage because $\langle T \rangle^4 \neq \langle T^4 \rangle$.

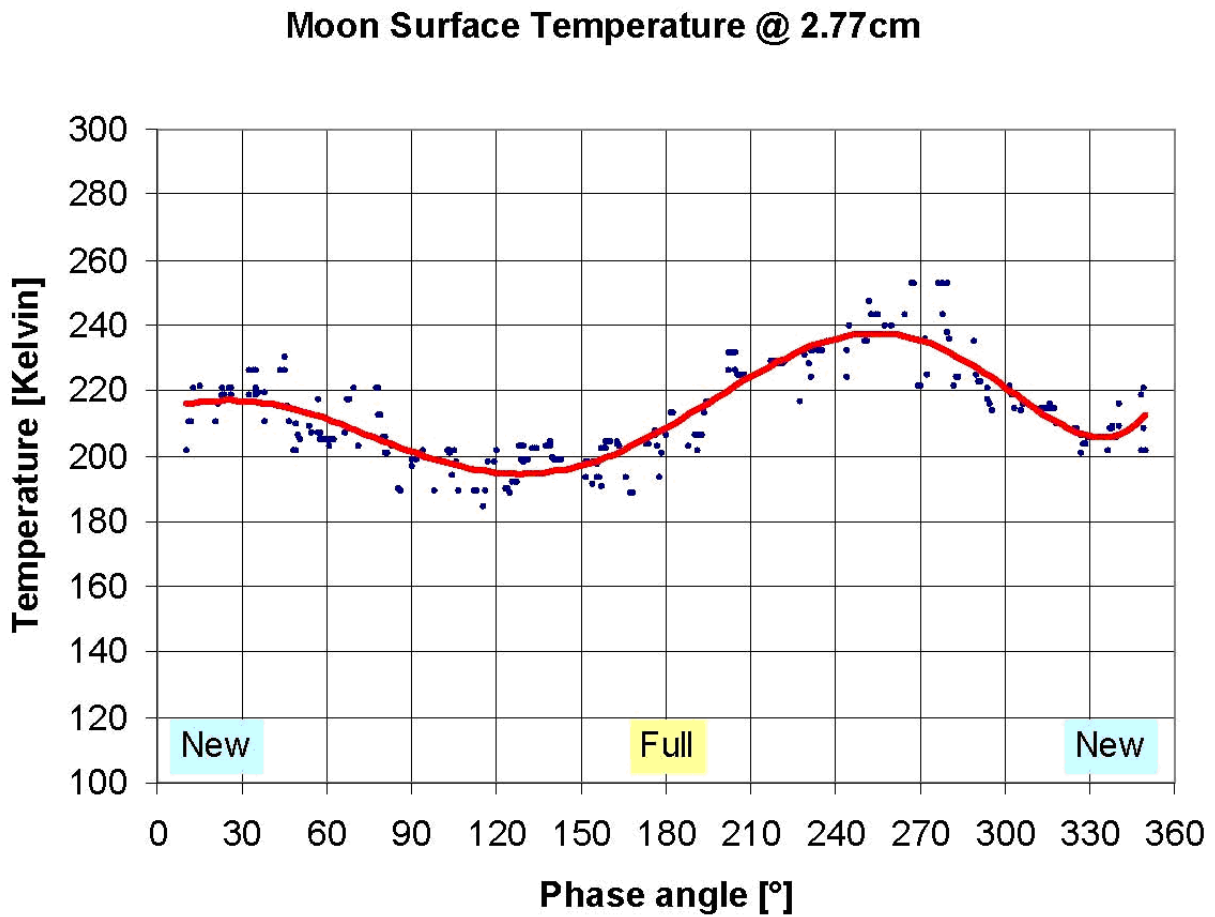

**Figure 5:** Moon's disk temperature at 2.77cm wavelength versus moon phase angle φ during two complete cycles from twice new moon via full moon to new moon again (adopted from Monstein, 2001).

Our Moon nearly satisfies the requirement of a planet without an atmosphere. It is well known that the Moon has no uniform temperature. There is not only a variation of the temperature from the lunar day to the lunar night, but also from the Moon equator to its poles.

Using Eq. (1.4) would provide $T_e \approx 270 \text{ K}$ when the albedo, $\alpha_M = 0.12$, and the emissivity, $\varepsilon_M = 1$ (black body), are considered. However, as illustrated in Figure 5, the mean disk temperature of the Moon observed at 2.77cm wavelength by Monstein (2001) is much lower than this equilibrium temperature.

Consequently, we have to consider that the Earth's surface temperature depends on longitude and latitude. As shown by Gerlich and Tscheuschner (2007, 2009), using Eq. (2.2) to planetary-averaged the temperature derived from a local radiation balance

$$T(\theta, \phi) = \left\{ \frac{(1 - \alpha(\theta, \phi)) F \cos(\Theta_0)}{\varepsilon(\theta, \phi) \sigma} \right\}^{\frac{1}{4}} \qquad (4.15)$$

leads to a value of

$$\langle T \rangle = \frac{2^{\frac{3}{2}}}{5} T_e \approx 144 \text{ K} \qquad (4.16)$$

for a non-rotating Earth, if $\alpha(\Theta_0, \theta, \phi) = \alpha_E$ and $\varepsilon(\theta, \phi) = \varepsilon_E = 1$ are assumed. Smith confirmed this value. His value (on page 7 of his paper) for a rotating Earth of about $\langle T \rangle = 252 \text{ K}$ or less is based on a numerical solution that needs confirmation.

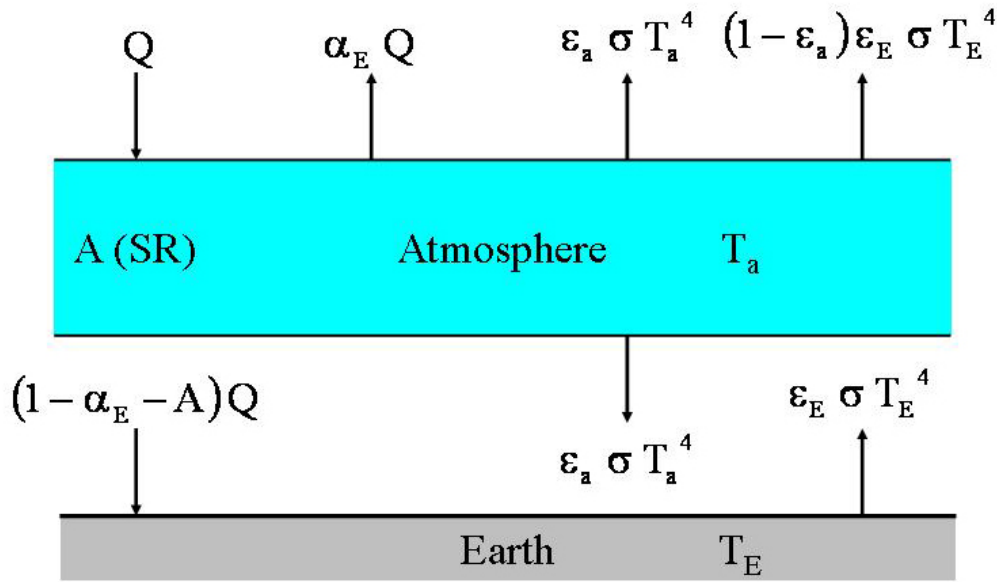

**Figure 6:** Two-layer model of the radiative equilibrium (adopted from Liou, 2002).

## 5. A two-layer model of radiative equilibrium

In his paper Smith also discussed the infrared absorption in the atmosphere. As illustrated in Figure 6, such a discussion leads to a two-layer model of radiative equilibrium. Inserting uniform temperatures for the atmosphere, $T_a$, and the Earth's surface, $T_E$ provides

*Top of the atmosphere:*

$$(1-\alpha_E)Q - \varepsilon_a \sigma T_a^4 - (1-\varepsilon_a)\varepsilon_E \sigma T_E^4 = 0 \quad , \tag{5.1}$$

*Earth's surface:*

$$(1-\alpha_E - A)Q + \varepsilon_E \varepsilon_a \sigma T_a^4 - \varepsilon_E \sigma T_E^4 = 0 \quad . \tag{5.2}$$

Here, $A$ is the atmospheric absorptivity in the range of solar radiation, and $Q = S/4$ to have the same input like in the case of the thought experiment of radiative equilibrium. The subscript a characterizes the values for the atmosphere and, again, the subscript E the values for the Earth's surface. Note that reflection of infrared radiation at the Earth's surface is included, but scattering of infrared radiation in the atmosphere is ignored. All properties are considered as uniform, too.

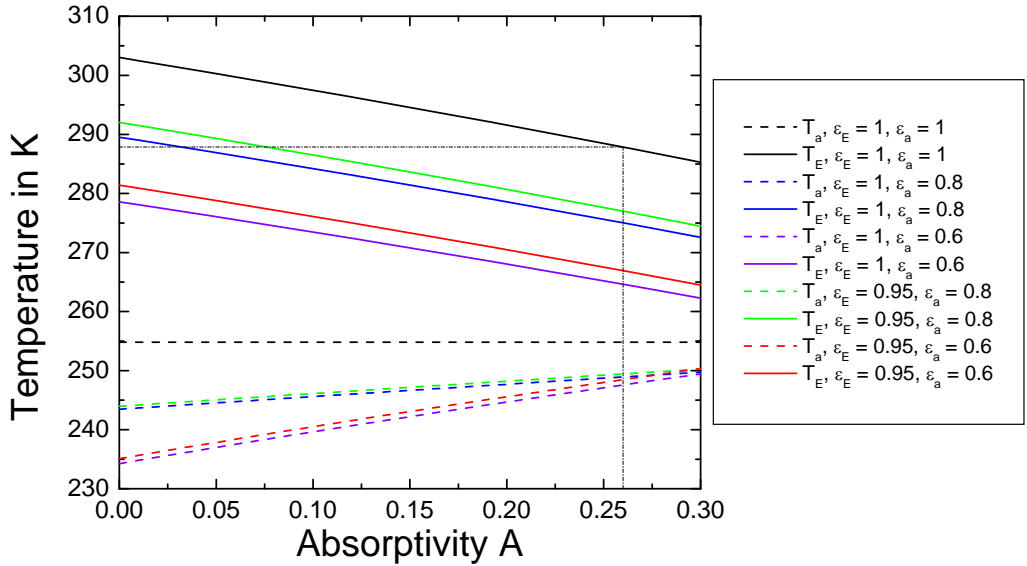

**Figure 7:** Uniform temperatures for the Earth's surface and the atmosphere provided by the two-layer model of radiative equilibrium versus absorptivity A.

The solution of this two-layer model of radiative equilibrium is given by

$$T_a = \left\{ \frac{\left(A + \varepsilon_a \left(1 - \alpha_E - A\right)\right) S}{4 \varepsilon_a \sigma \left(1 + \varepsilon_E \left(1 - \varepsilon_a\right)\right)} \right\}^{\frac{1}{4}} \tag{5.3}$$

and

$$T_E = \left\{ \frac{\left(\left(1 + \varepsilon_E\right)\left(1 - \alpha_E\right) - A\right) S}{4 \varepsilon_E \sigma \left(1 + \varepsilon_E \left(1 - \varepsilon_a\right)\right)} \right\}^{\frac{1}{4}}. \tag{5.4}$$

This pair of equation is non-linear with some coupling terms (Liou, 2002). The choice of $\varepsilon_E = 1$ leads to that of Liou (2002). It is obvious that $T_E$ and $T_a$ are dependent on the emissivity values of the Earth and the atmosphere, the absorption and the planetary albedo. Figure 7 illustrates results obtained for $\alpha_E \approx 0.30$ and some combinations of $\varepsilon_E$ and $\varepsilon_a$, where the absorptivity, $A$, is ranging from zero to 0.3. Assuming, for instance, that the atmosphere acts as blackbody emitter leads to an atmospheric temperature of about $T_a \cong 255\,K$ which is independent of the absorptivity (see Eq. (5.3) for $\varepsilon_a = 1$). Considering, in addition, the Earth as a blackbody emitter provides a surface temperature of about $T_E \cong 304\,K$ if the absorptivity is assumed to be zero. For the purpose of comparison: Smith obtained for a fully absorbing atmospheric layer $T_E = \sqrt[4]{2}\,T_{eff} \approx 303\,K$. Since the global average of temperatures observed in the close vicinity of the Earth's surface corresponds to $\langle T \rangle \approx 288\,K$, this value of $T_E \cong 304\,K$ suggests that the absorption of solar radiation in the atmosphere causes a decrease of the Earth's surface temperature (the atmospheric temperature increases concurrently). If we additionally assume that also the Earth acts as a blackbody emitter, we will obtain a temperature value for the Earth's surface of $T_e = 288\,K$ for an absorptivity of $A \cong 0.26$. This value is close to that estimated by Trenberth et al. (2009), namely $A \cong 0.23$ (see also Figure 8). However, as shown in Figure 7 other combinations of $\varepsilon_E$ and $\varepsilon_a$ provide different results. Even though that now $T_a$ grows with an increasing absorptivity, the decrease of $T_E$ is nearly as strong as the increase of $T_a$. Thus, we may conclude that the two-layer model of radiative equilibrium is, in principle, able to provide any pair of results for $T_E$ and $T_a$ we would like. This means that this two-layer model of radiative equilibrium is unsuitable for explaining the so-called greenhouse effect. Additionally, we must be aware that this simple consideration does not include any well known strong wavelength and pressure dependency of emissivity and absorption. Furthermore, the radiation flux balance at the Earth's surface has to be replaced by an energy flux balance,

$$\left(1 - \alpha_E - A\right) Q + \varepsilon_E \varepsilon_a \sigma T_a^4 - \varepsilon_E \sigma T_E^4 - H - E = 0 \tag{5.5}$$

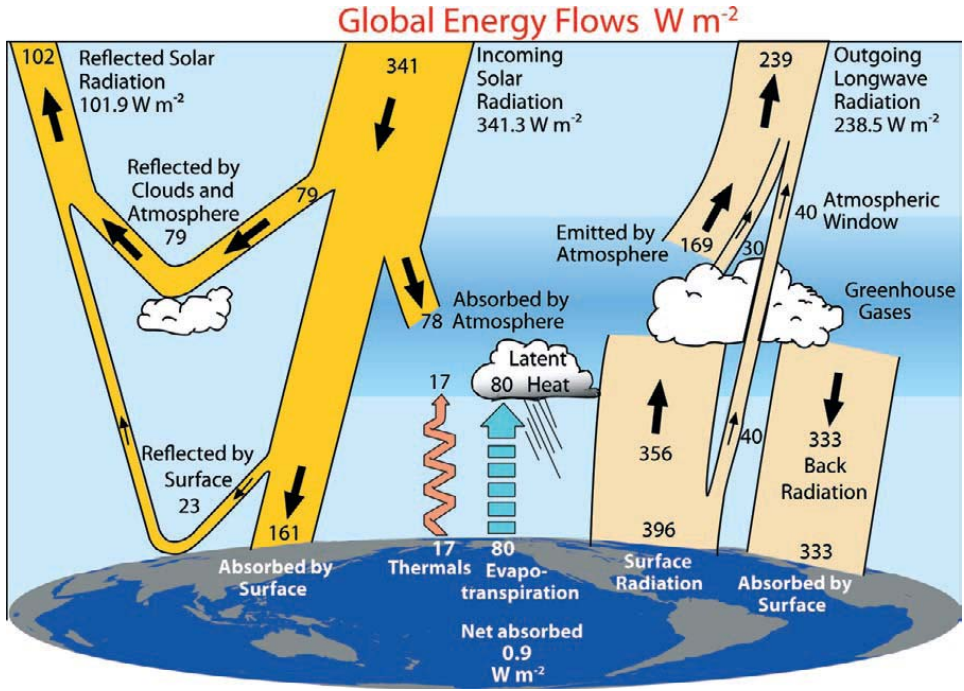

**Figure 8:** The global annual mean Earth's energy budget for the Mar 2000 to May 2004 period (W m$^{-2}$). The broad arrows indicate the schematic flow of energy in proportion to their importance (adopted from Trenberth et al., 2009).

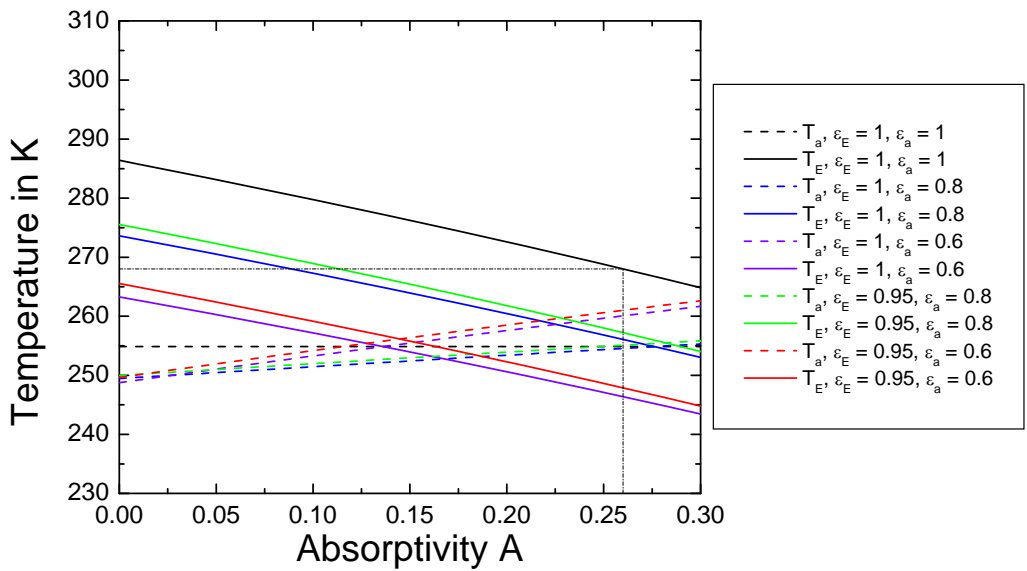

**Figure 9:** As in Figure 7, but the radiation flux balance at the Earth's surface is replaced by an energy flux balance including the fluxes of sensible and latent heat as suggested by Trenberth et al. (2009).

to include the fluxes of sensible heat, H, and latent heat, E. These two fluxes are not negligible (see Figure 8). In doing so, one obtains

$$T_a = \left\{ \frac{\left(A + \varepsilon_a\left(1 - \alpha_E - A\right)\right)\frac{S}{4} + \left(1 - \varepsilon_a\right)\left(H + E\right)}{\varepsilon_a \sigma \left(1 + \varepsilon_E\left(1 - \varepsilon_a\right)\right)} \right\}^{\frac{1}{4}} \tag{5.6}$$

and

$$T_E = \left\{ \frac{\left(\left(1 + \varepsilon_E\right)\left(1 - \alpha_E\right) - A\right)\frac{S}{4} - H - E}{\varepsilon_E \sigma \left(1 + \varepsilon_E\left(1 - \varepsilon_a\right)\right)} \right\}^{\frac{1}{4}}. \tag{5.7}$$

Results provided by this pair of equations are illustrated in Figure 9. This figure is based on the same combination of data as used in Figure 7. In accord with Trenberth et al. (2009) the fluxes of sensible and latent heat, $H = 17 \text{ W m}^{-2}$ and $E = 80 \text{ W m}^{-2}$, are considered. As illustrated, including the fluxes of sensible and latent heat leads to notably lower temperatures at the Earth's surface. In the case of $\varepsilon_E = 1.0$ and $\varepsilon_a = 0.6$ the surface temperature would be lower than the temperature of the radiative equilibrium of $T_e \approx 255 \text{ K}$, provided by Eq. (1.4), for $A \geq 0.15$. Thus, two-layer-models of radiative equilibrium are rather useless for proving the so-called $CO_2$ greenhouse effect. Smith's attempt to refute the criticism of Gerlich and Tscheuschner is, therefore, rather fruitless.

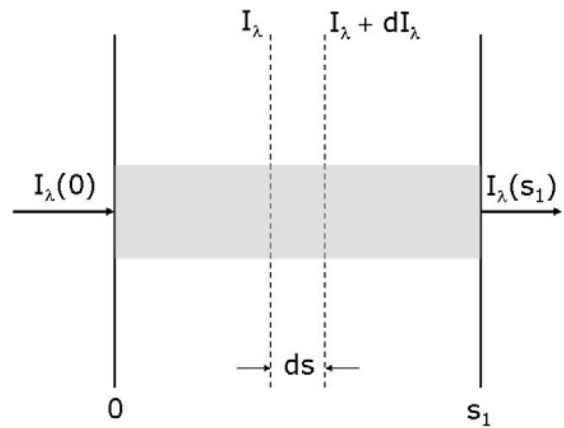

**Figure 10:** Depletion of the radiant intensity in traversing an extinction medium (adopted from Liou, 2002).

## 6. The radiative transfer equation

In an attempt to analyze how the absorption of solar and terrestrial (infrared) radiation and the emission of infrared radiation can contribute to the atmospheric budget of internal energy (alternatively enthalpy) as described by Eq. (3.17) it is indispensable to consider the radiative transfer equation (RTE). Its general form reads (e.g., Chandrasekhar, 1960; Liou, 2002)

$$\frac{1}{\rho \beta_\lambda} \frac{dI_\lambda}{ds} = -I_\lambda + J_\lambda \quad . \tag{6.1}$$

Here, $\beta_\lambda$ is the mass extinction cross section for radiation of the wavelength $\lambda$ (characterized by the subscript $\lambda$), $I_\lambda$ is the monochromatic intensity of radiation, $ds$ is the thickness of the layer crossing by the pencil of radiation (see Figure 10), and $J_\lambda$ is the monochromatic source function. Equation (6.1) describes two concurrent processes, namely attenuation of radiation by absorption and scattering on the one hand and emission of radiation by the constituents of the atmosphere plus multiple scattering of radiation from all other directions into this pencil of radiation on the other hand. The formal solution of Eq. (6.1) is given by (e.g., Chandrasekhar, 1960; Liou, 2002)

$$I_\lambda(s_1) = I_\lambda(0) \exp(-\tau_\lambda(s_1, 0)) - \int_0^{s_1} J_\lambda(T(s)) \exp(-\tau_\lambda(s_1, s)) d\tau_\lambda(s_1, s) \tag{6.2}$$

Here, $\tau_\lambda(s_1, s)$ is the monochromatic optical thickness defined by

$$\tau_\lambda(s_1, s) = \int_s^{s_1} \rho \beta_\lambda \, ds' \tag{6.3}$$

or

$$d\tau_\lambda(s_1, s) = -\rho \beta_\lambda \, ds \quad . \tag{6.4}$$

Equation (6.2) results in (a) Beer-Bouguer-Lamberts law by assuming only extinction due to absorption and/or scattering,

$$I_\lambda(s_1) = I_\lambda(0) \exp(-\tau_\lambda(s_1, 0)) \quad , \tag{6.5}$$

and (b) the Schuster-Schwarzschild equation by considering a non-scattering medium and local thermodynamic equilibrium (approximately fulfilled in the atmosphere up to 60 km above the Earth's surface, e.g., Lenoble, 1993; Liou, 2002) so that the source function may be replaced by the Planck function given in the wavelength domain $[0, \infty]$

$$B(\lambda, T) = \frac{2\,h\,c^2}{\lambda^5 \left\{ \exp\left(\dfrac{h\,c}{\lambda\,k\,T}\right) - 1 \right\}} \quad . \tag{6.6}$$

where now $\beta_\lambda$ becomes the mass absorption coefficient. In Eq. (6.6), $h = 6.626 \cdot 10^{-34}$ J s is the Planck constant, $c = 2.998 \cdot 10^8$ m s$^{-1}$ is the velocity of light in vacuum, and $k = 1.3806 \cdot 10^{-23}$ J K$^{-1}$ is the Boltzmann constant. Inserting the Planck function into Eq. (6.2) yields

$$I_\lambda(s_1) = I_\lambda(0)\exp(-\tau_\lambda(s_1, 0)) - \int_0^{s_1} B_\lambda(T(s))\exp(-\tau_\lambda(s_1, s))\,d\tau_\lambda(s_1, s) \quad . \tag{6.7}$$

This equation seems to be well appropriate when the infrared radiation emitted by both the Earth and the constituents of the atmosphere is considered as long as local thermodynamic equilibrium is guaranteed. Obviously, the first term of the right-hand side of Eq. (6.7) describes the absorption of infrared radiation along the path $[0, s_1]$ and the second term the emission of infrared radiation along the path $[0, s_1]$ performed simultaneously. This equation has to be numerically integrated for practical purposes by considering all important absorption lines/bands. Note, however, that only the divergence of the total radiation flux density expressed by $\nabla \cdot \overline{\mathbf{R}}$ can contribute to a change of internal energy (see Eq. (3.17)).

## 7. Summary and Conclusions

It was shown that Smith's formulations of planetary averages are rather inappropriate and inconsistently used. Introducing for different averages to perform the respective calculations makes no sense because such calculations can be realized without them. Only the definition Eq. (2.2) is required to derive formula (4.16).

In research on turbulence it is indispensable and widely recognized that all governing equations for turbulent systems have been derived in an entirely consistent manner. Changing averaging procedures from one quantity to another as can be found in Smith's paper is neither advantageous nor reasonable in theoretical studies.

Smith's discussion of the infrared absorption in the atmosphere was scrutinized and evaluated. It was shown that his attempt to refute the criticism of Gerlich and Tscheuschner (2007, 2009) on the so-called greenhouse effect is rather fruitless. To study how the absorption of solar and terrestrial (infrared) radiation and the emission of infrared radiation can contribute to the atmospheric budget of internal energy (alternatively enthalpy) it is indispensable to consider the radiative transfer equation along with the first and second laws of thermodynamics as already pointed out by Vilhelm Bjerkness (1904).

**Appendix A**

The total flux (also called the radiant power) of solar radiation, $E_{S\downarrow}$, reaching the surface of an Earth without an atmosphere is given by

$$E_{S\downarrow} = \int_A \mathbf{F} \cdot \mathbf{n}(\mathbf{r}) \, dA(\mathbf{r}) \quad . \tag{A1}$$

Here, $\mathbf{F}$ is the flux density of solar radiation also called the solar irradiance and F is its magnitude, A is the radiation-exposed surface of the Earth, $dA(\mathbf{r})$ the corresponding differential surface element, and $\mathbf{n}(\mathbf{r})$ is the inward pointing unit vector perpendicular to $dA(\mathbf{r})$, where the direction of $\mathbf{n}(\mathbf{r})$ is chosen in such a sense that $\mathbf{F} \cdot \mathbf{n}(\mathbf{r}) \geq 0$ is counted positive. This unit vector is also called the unit normal. Its origin is the tip of the position vector $\mathbf{r}$ with which the location of $dA(\mathbf{r})$ on the Earth's surface is described. The angle between $\mathbf{F}$ and $\mathbf{n}(\mathbf{r})$ is the local zenith angle of the Sun's center, $\Theta_0(\mathbf{r})$. The scalar product $\mathbf{F} \cdot \mathbf{n}(\mathbf{r})$ that describes the solar radiation reaching the horizontal surface element, is given by $\mathbf{F} \cdot \mathbf{n}(\mathbf{r}) = F \cos \Theta_0(\mathbf{r})$.

From now on we simply consider the Earth as a sphere; and its center as the origin of the position vector so that $\mathbf{r} = \mathbf{r}_E$. Thus, $\mathbf{n}(\mathbf{r}_E)$ and $\mathbf{r}_E$ are collinear vectors having opposite directions. Furthermore, $A = r_E^2 \, \Omega$ with $\Omega = 2\pi$ is the solid angle of a half sphere, and $dA = r_E^2 \, d\Omega = r_E^2 \sin\theta \, d\theta \, d\phi$ (see Figure 3). Thus, the radiant power can, therefore, be written as

$$E_{S\downarrow} = r_E^2 \int_\Omega F \cos \Theta_0 \, d\Omega = r_E^2 \int_0^{2\pi} \int_0^{\pi/2} F \cos \Theta_0 \sin\theta \, d\theta \, d\phi \quad . \tag{A2}$$

Since, however, a portion of $F \cos \Theta_0(\mathbf{r})$ is not absorbed by the skin of the Earth's surface because it is diffusely reflected we have to introduce the local albedo, $\alpha(\Theta_0, \theta, \phi)$, into this equation. The total flux of solar radiation that is absorbed by the Earth's skin, $F_{S\downarrow}$, is then given by

$$F_{S\downarrow} = r_E^2 \int_\Omega F\left(1 - \alpha(\Theta_0, \theta, \phi)\right) \cos \Theta_0 \, d\Omega = r_E^2 \int_0^{2\pi} \int_0^{\pi/2} F\left(1 - \alpha(\Theta_0, \theta, \phi)\right) \cos \Theta_0 \sin\theta \, d\theta \, d\phi \quad , \tag{A3}$$

Equation (4.1) is identical with Eq. (A3). Note that the local albedo also depends on $\Theta_0(\mathbf{r})$.

The total flux of infrared radiation emitted by the Earth's surface is given by

$$F_{IR\uparrow} = \int_A \mathbf{E}(\mathbf{r}_E) \cdot \mathbf{n}(\mathbf{r}_E) \, dA(\mathbf{r}_E) \quad . \tag{A4}$$

Here, $\mathbf{E}(\mathbf{r}_E) = \varepsilon(\mathbf{r}_E) \sigma T^4(\mathbf{r}_E) \mathbf{e}_r$ is the flux density of infrared radiation at the location $\mathbf{r}_E$, and $E(\mathbf{r}_E) = \varepsilon(\mathbf{r}_E) \sigma T^4(\mathbf{r}_E)$ is its magnitude. The emission of radiant energy by a small surface element into the adjacent half space is considered as isotropic, as required by the derivation of the power law of Stefan (1879) and Boltzmann (1884). The unit vector $\mathbf{e}_r = \mathbf{r}_E / |\mathbf{r}_E|$ points to the zenith of this adjacent half space. Note that $\mathbf{e}_r$ and $\mathbf{n}(\mathbf{r}_E)$ are collinear unit vectors having opposite directions. Thus, we have $\mathbf{e}_r \cdot \mathbf{n}(\mathbf{r}_E) = -1$. This only means that $F_{S\downarrow}$ is counted positive, and $F_{IR\uparrow}$ is counted negative so that the planetary radiative equilibrium reads: $F_{S\downarrow} - F_{IR\uparrow} = 0$ or $F_{S\downarrow} = F_{IR\uparrow}$. The radiant power in the infrared range is then given by

$$F_{IR\uparrow} = r_E^2 \int_\Omega E(\theta, \phi) d\Omega = r_E^2 \int_0^{2\pi} \int_0^\pi E(\theta, \phi) \sin\theta \, d\theta \, d\phi \quad , \tag{A5}$$

where $\Omega = 4\pi$ is the solid angle for the entire planet, and $E(\theta, \phi) = \varepsilon(\theta, \phi) \sigma T^4(\theta, \phi)$. Equation (4.11) is identical with Eq. (A5).